\documentclass[letterpaper, 10 pt, conference]{ieeeconf}
\IEEEoverridecommandlockouts
\usepackage{graphics} 
\usepackage{amsmath} 
\usepackage{amssymb}  
\usepackage{physics}
\usepackage{cite}
\usepackage{esdiff}
\usepackage{xcolor}
\usepackage{siunitx}
\usepackage{graphicx}
\usepackage[font={footnotesize}]{caption}
\usepackage{float}
\usepackage{subfig}
\usepackage{flushend}
\usepackage{amsthm}
\theoremstyle{definition}\newtheorem{definition}{Definition}
\graphicspath{ {./img/} }
\def\BibTeX{{\rm B\kern-.05em{\sc i\kern-.025em b}\kern-.08em
    T\kern-.1667em\lower.7ex\hbox{E}\kern-.125emX}}    

\title{\LARGE \bf An Optimal Control Framework for Influencing Human Driving Behavior in Mixed-Autonomy Traffic}

\author{Anirudh Chari$^{1}$, Rui Chen$^{2}$, Jaskaran Grover$^{3}$, Changliu Liu$^{2}$
\thanks{$^{1}$A. Chari is with the Illinois Mathematics and Science Academy in Aurora, IL, USA. (\tt{achari@imsa.edu})}%
\thanks{$^{2}$R. Chen and C. Liu are with the Robotics Institute at Carnegie Mellon University in Pittsburgh, PA, USA. (\tt{\{ruic3,cliu6\}@andrew.cmu.edu})}%
\thanks{$^{3}$J. Grover is with Amazon Robotics in North Reading, MA, USA. (\tt{jessygro@amazon.com})}%
\thanks{This work is partially supported by the National Science Foundation, Grant No. 2144489.}%
}

\begin{document}

\maketitle
\thispagestyle{empty}
\pagestyle{empty}

\begin{abstract}
As autonomous vehicles (AVs) become increasingly prevalent, their interaction with human drivers presents a critical challenge.
Current AVs lack social awareness, causing behavior that is often awkward or unsafe.
To combat this, social AVs, which are proactive rather than reactive in their behavior, have been explored in recent years.
With knowledge of robot-human interaction dynamics, a social AV can influence a human driver to exhibit desired behaviors by strategically altering its own behaviors.
In this paper, we present a novel framework for achieving human influence.
The foundation of our framework lies in an innovative use of control barrier functions to formulate the desired objectives of influence as constraints in an optimal control problem.
The computed controls gradually push the system state toward satisfaction of the objectives, e.g. slowing the human down to some desired speed.
We demonstrate the proposed framework's feasibility in a variety of scenarios related to car-following and lane changes, including multi-robot and multi-human configurations.
In two case studies, we validate the framework's effectiveness when applied to the problems of traffic flow optimization and aggressive behavior mitigation.
Given these results, the main contribution of our framework is its versatility in a wide spectrum of influence objectives and mixed-autonomy configurations.
\end{abstract}


\section{Introduction}\label{intro}
Autonomous vehicles (AVs) are increasingly popular on today's roadways, largely as a result of great strides in perception, planning, and control algorithms in recent years \cite{ppc}.
However, these systems remain far from perfect, particularly because current AVs struggle to deal with human drivers.
Many existing works treat human drivers as dynamic obstacles rather than decision-making agents \cite{safety1}, and this assumption often yields awkward or unsafe behavior.
Addressing this problem is crucial, as AVs and human drivers will likely share the roads for many years to come \cite{adoption}.

Recently, the idea of a \textit{social} AV has emerged as a potential solution to this problem, with human-AV social interactions being studied extensively \cite{social}.
The social AV is proactive rather than reactive: it leverages models of human behavior to inform its own decision-making.
In particular, a social AV might strategically alter its own behavior so as to \textit{influence} a human driver to exhibit some desired behavior.
To accomplish this, the AV must exploit models not only of the human's driving behavior but also of the change in the human's behavior with respect to its own behavior.
The concept of social influence has numerous applications in mixed-autonomy settings.
In congested traffic, AVs can orchestrate cooperative merging and lane-changing maneuvers, alleviating traffic bottlenecks. 
In emergency situations, AVs can guide human drivers to take evasive actions that mitigate collision risks.
Human influence for general human-robot collaboration is studied in \cite{ravi}, and applications to traffic flow optimization have been widely explored \cite{flow1,flow2}.
Furthermore, a single-agent control framework is provided in \cite{dorsa1}, and \cite{probe,simult1} provide frameworks for single-agent influence with simultaneous probing, i.e. learning human driving parameters by observing behavior.

More generally, the concept of a socially \textit{aware} AV has been studied extensively.
Awareness is a prerequisite for influence: a social AV must first understand human interactions before attempting to alter human behavior.
While human influence is a very active procedure, awareness is more passive.
The socially aware AV is cognizant of its effects on human behaviors and it may consider these effects during planning, but it does not necessarily seek to apply this knowledge to altering these behaviors.
Social awareness functions as a significant first step toward full human-AV social interaction in mixed-autonomy settings, especially given the extensive work done in learning human driving behavior \cite{lanechange3,rl1}.
Some studies have achieved superior AV control by considering human car-following \cite{follow} and lane-changing \cite{lc1} behaviors.
These ideas have also been employed in designing AV control frameworks \cite{when}.

\begin{figure}[t]
    \centering
    \includegraphics[width=\columnwidth]{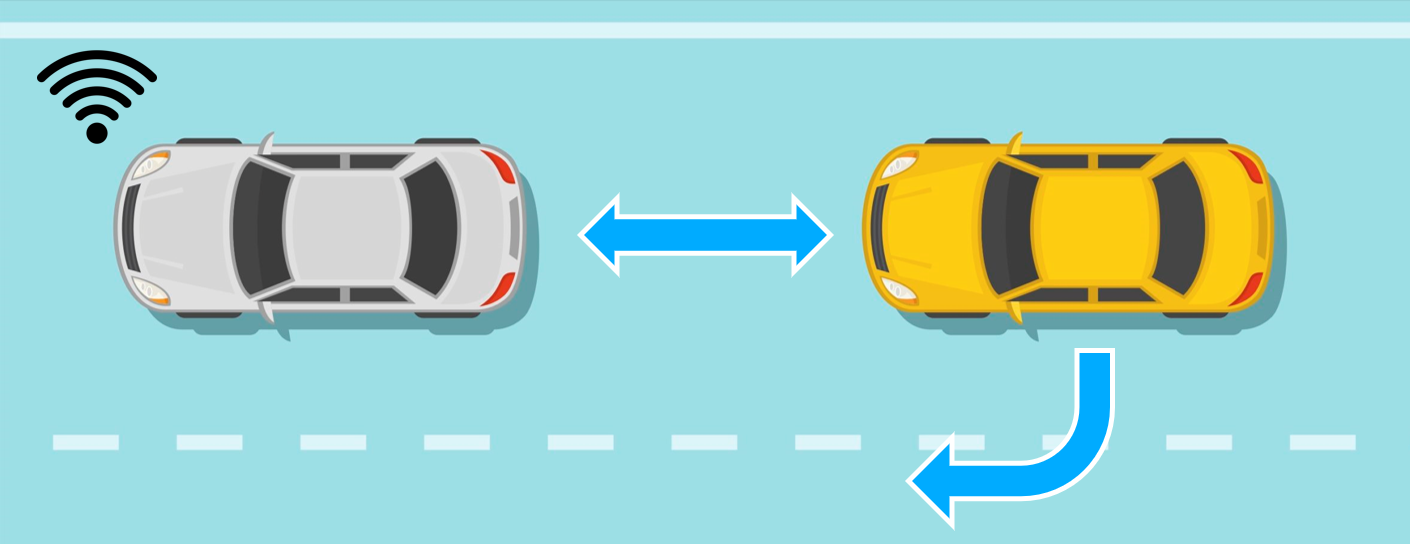}
    \caption{A robot car is being tailgated by a human-driven car. How can the robot influence the human to increase following distance or to change lanes?}
    \label{fig:first-pic}
\end{figure}

Control barrier functions (CBFs) are a popular tool for safe AV control.
In particular, CBFs have been employed in problems of cooperative AV merging \cite{dolan-merge,cass-merge} and lane-changing \cite{lc-cbf}.
For a thorough overview of CBFs and their applications, see \cite{cbfs}.
In comparison to existing methods of human influence that traditionally use game-theoretic formulations \cite{game-theory,ilq-games}, a CBF-based approach offers guarantees of safety and convergence, resulting in more robust and predictable policies.

To the best of the authors' knowledge, the current literature on human influence is restricted either to localized objectives (e.g. following, merging) or to single-agent configurations.
In this work, we aim to fill this gap.
\textbf{Our main contributions are: (a) an optimal control framework for influencing human driving behavior using CBFs, with applicability to (b) various objectives of influence and (c) multi-robot and multi-human scenarios.}
Section \ref{probform} formulates the problem of human influence.
Section \ref{method} presents our framework.
Section \ref{experiments} discusses the results of principal experiments using our framework.
Section \ref{cases} discusses the results of two case studies involving our framework.
Finally, Section \ref{conclusion} concludes our study and presents future directions.

\section{Problem Formulation}\label{probform}
We are given a group of $m$ human-driven cars, $n$ robot cars, and $q$ non-interactive background cars that travel along a multi-lane highway with $L$ lanes.
While human-driven cars react to the behavior of robot cars, background cars do not.
Our objective is to compute robot car controls $\boldsymbol{u}$ at each timestep that influence the human-driven cars to exhibit some desired behaviors and minimize some cost function $J(\boldsymbol{u})$.

In a state space $\mathcal{X}$, let $\boldsymbol{p}_H \in \mathbb{R}^m$ denote the human-driven car positions relative to some fixed starting point along their respective lanes, $\boldsymbol{v}_H \in \mathbb{R}^m$ their velocities in their lanes, and $\boldsymbol{a}_H \in \mathbb{R}^m$ their accelerations in their lanes. 
Similarly, let $\boldsymbol{p}_{R} \in \mathbb{R}^n$ denote positions of the robot cars along their lanes, $\boldsymbol{v}_{R} \in \mathbb{R}^n$ their velocities, and $\boldsymbol{a}_{R} \in \mathbb{R}^n$ their accelerations, and let $\boldsymbol{p}_{B} \in \mathbb{R}^q$ denote positions of the background cars along their lanes, $\boldsymbol{v}_{B} \in \mathbb{R}^q$ their velocities, and $\boldsymbol{a}_{B} \in \mathbb{R}^q$ their accelerations.
Let $p_{Hi}$ denote the position of the $i$-th human-driven car $H_i$, $p_{Rj}$ the position of the $j$-th robot car $R_j$, and $p_{Bk}$ the position of the $k$-th background car $B_k$.
Let $l_{Hi} \in \{1, ..., L\}$ denote the current lane of human-driven car $H_i$, and represent the current lanes of all human-driven cars as $\boldsymbol{l}_H := (l_{H1}, ..., l_{Hm})$.
Define $\boldsymbol{l}_R$ for the current lanes of the robot cars and $\boldsymbol{l}_B$ for the current lanes of the background cars in a similar fashion.
We will assume that robot cars and background cars maintain their lanes.

Let us assume that human-driven cars follow double integrator dynamics that can be represented as
\begin{align}
    \Dot{\boldsymbol{p}}_{H} = \boldsymbol{v}_{H},~ \Dot{\boldsymbol{v}}_{H} = \boldsymbol{a}_{H} = \boldsymbol{f}(\boldsymbol{x})
\end{align}
where $\boldsymbol{x}$ is the system state and each $f_i(\boldsymbol{x})$ has some known parameters $\boldsymbol{\theta}_i$ that define human driver $i$'s behavior \cite{idm}. 
Let us assume that the robot cars are acceleration-controlled. 
Their dynamics can be posed as
\begin{align}
 \Dot{\boldsymbol{p}}_{R} = \boldsymbol{v}_{R}, ~\Dot{\boldsymbol{v}}_{R} = \boldsymbol{a}_{R} \equiv \boldsymbol{u}.
\end{align}
Let us assume that the background cars travel with constant velocities. 
This is a reasonable assumption because, in the absence of external influence, cars generally maintain their speed in highway settings.
Then we have
\begin{align}
    \dot{\boldsymbol{p}}_B = \boldsymbol{v}_B, ~\dot{\boldsymbol{v}}_B = \boldsymbol{a}_B = \boldsymbol{0}.
\end{align}

The human-driven cars have lane-change controls $\boldsymbol{d}_H \in \{-1, 0, 1\}^m$. 
Lane changing is a discrete event; lanes are updated at each timestep as
\begin{equation}
    \boldsymbol{l}_H (t + \Delta t) = \boldsymbol{l}_H (t) + \boldsymbol{d}_H.
\end{equation}
This definition encodes the actions \{left, stay, right\}.
For human-driven car $H_i$ we have
\begin{equation}\label{eq:dhi}
    d_{Hi} = \pm (z^S_i(\boldsymbol{x}) \geq 0 \hspace{2pt} \wedge \hspace{2pt} z^I_i(\boldsymbol{x}) \geq 0)
\end{equation}
where $z^S_i(\boldsymbol{x})$ is a lane-change safety function, $z^I_i(\boldsymbol{x})$ is a lane-change incentive function, and both depend again on parameters $\boldsymbol{\theta}_i$.
For example, $z^S_i(\boldsymbol{x})$ could be a function of the space available between cars in the adjacent lane, and $z^I_i(\boldsymbol{x})$ could be a function of the difference in velocity or acceleration between the human's current lane and the adjacent lane.
The $\pm$ symbol exists here to denote that a lane change may occur in either direction.

We denote the human-driven cars' collective state as $\boldsymbol{x}_H := (\boldsymbol{p}_H, \boldsymbol{v}_H, \boldsymbol{l}_H)$.
We denote the robot cars' collective state as $\boldsymbol{x}_R := (\boldsymbol{p}_R, \boldsymbol{v}_R, \boldsymbol{l}_R)$.
We denote the background cars' collective state as $\boldsymbol{x}_B := (\boldsymbol{p}_B, \boldsymbol{v}_B, \boldsymbol{l}_B)$.
The total state of our system is denoted as $\boldsymbol{x} := (\boldsymbol{x}_H, \boldsymbol{x}_R, \boldsymbol{x}_B)$.

We can formulate an arbitrary objective of influence as a constraint function $\psi : \mathcal{X} \rightarrow \mathbb{R}$ such that $\psi \geq 0$ implies that the objective is satisfied.
For example, we can express influencing an upper bound $v_{max}$ on human-driven car $H_i$'s velocity as $\psi =  v_{max} - v_{Hi}$.
Because lane changes are discrete events, we will utilize \eqref{eq:dhi} to derive constraints that influence lane changes, rather than formulating constraints directly in terms of $\boldsymbol{d}_H$.
For example, if $z^I_i$ is a function of $H_i$'s velocity, then our constraint will be directly on $H_i$'s velocity.
For $N$ simultaneous influence objectives, we can derive constraints $\boldsymbol{\psi} := (\psi_1, ..., \psi_N)$.
Then given some cost function $J(\boldsymbol{u})$, we can pose the following problem to solve for optimal robot controls $\boldsymbol{u}^*$ that achieve human influence:
\begin{equation}\label{eq:probform_opt}
    \boldsymbol{u}^* = \operatorname*{argmin}_{\boldsymbol{u}} \hspace{5pt} J(\boldsymbol{u}) ~~~ \text{s.t.} ~~ \boldsymbol{\psi} \geq \boldsymbol{0}
\end{equation}
In the next section, we further derive a control constraint that yields satisfaction of the state constraint $\boldsymbol{\psi} \geq \boldsymbol{0}$.

\section{Optimal Control Framework}\label{method}
In this section, we present the optimal control framework in detail.
We first describe the intuition motivating the framework's design.
We then review and define some key properties of time derivatives.
Next, we outline the procedure by which constraints on the robot controls are derived.
Finally, we pose the general optimal control problem for achieving human influence.

\subsection{Preliminary: Control Barrier Functions}\label{cbf}
In \cite{herd}, CBFs are employed to compute controls for a group of \textit{dog robots} that herd a flock of \textit{sheep agents}.
The sheep aim to reach some goal location, and the dogs must prevent the sheep from breaching some spatial protected zone.
Crucially, the dog robots exploit knowledge of the dog-sheep interaction dynamics to achieve this objective.
Similarly, here we have a group of human-driven cars that exhibit some defined behavior, and a group of surrounding robot cars must influence this behavior with the objective of enforcing some abstract ``protected zone.''

We claim that CBFs are a useful tool for enforcing this protected zone.
Using CBFs, we define our protected zone as a barrier function $h : \mathcal{X} \rightarrow \mathbb{R}$ where $h \geq 0$ implies safety in our system, meaning satisfaction of our influence objective.
To push our system toward safety over time, we can enforce $\dot{h} + p_1 h \geq 0$, where $p_1$ is a design parameter and $p_1 > 0$.
To derive an explicit constraint on robot controls, we might compute further time derivatives of $h$ and express our constraint in terms of $\ddot{h}$ or $\dddot{h}$.
In general, an $N$-th order constraint is given as $(1 + p_1 s)(1 + p_2 s) \cdots (1 + p_N s) h \geq 0$ where $s$ is the differentiation operator.
We expand this to $h + \sum_{i=0}^{N} \alpha_i h^{(i)} \geq 0$ where $h^{(i)}$ denotes the $i$-th time derivative of $h$.
It remains that our system tends toward $h \geq 0$, and thus the protected zone is enforced.

\subsection{Preliminary: Time Derivative Properties}

In general, we can make use of the following discrete approximations for computing time derivatives:
\begin{equation}\label{eq:discrete-fi}
    \dot{f}_i = \diff{f_i}{t} \approx \frac{f_i - a_{Hi}}{\Delta t}
\end{equation}
\begin{equation}\label{eq:discrete-uj}
    \dot{u}_j = \diff{u_j}{t} \approx \frac{u_j - a_{Rj}}{\Delta t}
\end{equation}
where $\Delta t$ is the discretization timestep.

\begin{definition}[Direct influence]
We refer to a human-driven car $H_i$ as \textit{directly influenced} by a robot car $R_j$ if $f_i(\boldsymbol{x})$ is dependent on $\boldsymbol{x}_{Rj}$, i.e. $\nabla_{\boldsymbol{x}_{Rj}} f_i(\boldsymbol{x}) \neq \boldsymbol{0}$.
\end{definition}

Consider a human-driven car $H_i$ that is directly influenced by a robot car $R_j$.
Then when computing the time derivative of $f_i$, we have
\begin{align}\label{eq:fi-dot}
    \dot{f}_i (\boldsymbol{x}) &= \nabla f_i(\boldsymbol{x}(t))^T \dot{\boldsymbol{x}}(t) \nonumber \\
    &= \pdv{f_i}{p_{Hi}} \dot{p}_{Hi} + \pdv{f_i}{v_{Hi}} \dot{v}_{Hi} + \pdv{f}{p_{Rj}} \dot{p}_{Rj} + \pdv{f_i}{v_{Rj}} \dot{v}_{Rj} \nonumber \\
    &= \pdv{f_i}{p_{Hi}} v_{Hi} + \pdv{f_i}{v_{Hi}} a_{Hi} + \pdv{f_i}{p_{Rj}} v_{Rj} + \pdv{f_i}{v_{Rj}} u_j \nonumber \\
    &= \lambda_{ij} + \pdv{f_i}{v_{Rj}} u_j
\end{align}
where $\lambda_{ij} := \pdv{f_i}{p_{Hi}} v_{Hi} + \pdv{f_i}{v_{Hi}} a_{Hi} + \pdv{f_i}{p_{Rj}} v_{Rj}$.
Notice that this gives us an explicit relationship between human actions and robot controls.

\subsection{Constraint Derivation}
Let $y^{(c)}$ denote the $c$-th time derivative of $y$.
For our overall state constraint $\boldsymbol{\psi} \geq \boldsymbol{0}$, recall that each individual $c$-th order constraint is defined as $\psi_i (\boldsymbol{p}_{H}^{(c)}, \boldsymbol{p}_{R}^{(c)}, \boldsymbol{p}_{B}^{(c)})$ and encodes some desired high-level objective such that $\psi_i \geq 0$ iff the objective is satisfied.
Thus, $\psi_i$ can be treated as a barrier function $h$, as outlined in Section \ref{cbf}.

Note that not all terms in $\psi_i$ are necessarily nonzero, e.g. an objective may only involve a subset of the cars in $\boldsymbol{x}$.
Also note that we require the same relative degree for all terms, although this is sometimes aided by approximating higher-order terms using lower-order terms.

We compute $b$ time derivatives of $\psi_i$ to obtain
\begin{align}
    \psi_i^{(b)} (\boldsymbol{p}_{H}^{(b+c)}, \boldsymbol{p}_{R}^{(b+c)}) &:= \psi_i^h (\dot{\boldsymbol{f}}, \dot{\boldsymbol{u}}) \nonumber \\
    &\equiv \psi_i^h (\dot{\boldsymbol{f}}_{\lnot \varphi}, \dot{\boldsymbol{f}}_{\varphi}, \dot{\boldsymbol{u}})
\end{align}
where $\dot{\boldsymbol{f}}_{\varphi}$ is the set of all $\dot{f}_i$ such that human-driven car $H_i$ is directly influenced by some robot car $R_j$, and $\dot{\boldsymbol{f}}_{\lnot \varphi}$ is the set of all $\dot{f}_i$ such that human-driven car $H_i$ is \underline{not} directly influenced by any robot car $R_j$.
Notice that we no longer have a $\boldsymbol{p}_B$ term, since we assume constant velocities for background cars.

We first leverage \eqref{eq:fi-dot} to convert $\dot{\boldsymbol{f}}_{\varphi}$ terms to $\boldsymbol{u}$ terms, yielding the equivalent form $\psi_i^{(b)}(\dot{\boldsymbol{f}}_{\lnot \varphi}, \boldsymbol{u}, \dot{\boldsymbol{u}})$.
Then, we leverage \eqref{eq:discrete-fi} to convert $\dot{\boldsymbol{f}}_{\lnot \varphi}$ terms to $\boldsymbol{f}$ terms, and \eqref{eq:discrete-uj} to convert $\dot{\boldsymbol{u}}$ terms to $\boldsymbol{u}$ terms, yielding a more usable function $\psi_i^h(\boldsymbol{f}, \boldsymbol{u})$.
Finally, we can express our initial objective as the following CBF constraint:
\begin{equation}
    g_i(\boldsymbol{x}, \boldsymbol{f}, \boldsymbol{u}) := \psi_i^h + \sum_{j=0}^{b-1} \alpha_j \psi_i^{(j)} \geq 0
\end{equation}
Using CBFs, we have now effectively translated our original state constraint $\psi_i \geq 0$ into a control constraint $g_i \geq 0$.

\subsection{Optimization Problem}
We use the above process to derive $N$ linear constraints:
\begin{equation}\label{eq:g-def}
\boldsymbol{g}(\boldsymbol{x}, \boldsymbol{f}, \boldsymbol{u}) :=
    \begin{bmatrix}
        g_1(\boldsymbol{x}, \boldsymbol{f}, \boldsymbol{u}) \\
        \vdots \\
        g_N(\boldsymbol{x}, \boldsymbol{f}, \boldsymbol{u})
    \end{bmatrix}
\geq \boldsymbol{0}
\end{equation}
We can then adapt \eqref{eq:probform_opt} to pose the following optimization problem to solve for controls $\boldsymbol{u}^*$ that push the system toward $\boldsymbol{\psi} \geq \boldsymbol{0}$, i.e. satisfaction of the objectives of influence:
\begin{equation}\label{eq:big-opt}
    \boldsymbol{u}^* = \operatorname*{argmin}_{\boldsymbol{u}} \hspace{5pt} J(\boldsymbol{u}) ~~~ \text{s.t.} ~~ \boldsymbol{g}(\boldsymbol{x}, \boldsymbol{f}, \boldsymbol{u}) \geq \boldsymbol{0}
\end{equation}
The problem $\eqref{eq:big-opt}$ is feasible if \eqref{eq:g-def} is feasible (i.e. the half-spaces created by $g_1, ..., g_N$ intersect) and \eqref{eq:g-def} intersects with the control limits.
In this paper, we assume that \eqref{eq:big-opt} is always feasible.
To ensure such feasibility in practice, one must solve the synthesis problem, as introduced in \cite{zhao2023sos,chen2023sis}.

\section{Principal Experiments}\label{experiments}
In this section, we verify the feasibility of our framework under various objectives by simulating nine low-level scenarios.
We also provide an example constraint derivation for one scenario to illustrate the process.

\subsection{Human Behavior Models}
Here we specify the behavior models used for human driver control.
For longitudinal control (i.e., car-following), we define each $f_i(\boldsymbol{x})$ using the intelligent driver model (IDM) \cite{idm}.
Thus, we have
\begin{equation}
    f_i := a_i^{max} \left( 1 - \left( \frac{v_{Hi}}{v_{0i}} \right) ^4 - \left( \frac{s^*(v_{Hi}, \Delta v_{i})}{s_i} \right) ^2 \right)
\end{equation}
with
\begin{equation}
    s^*(v_{Hi}, \Delta v_{i}) := s_{0i} + v_{Hi} T + \frac{v_{Hi} \Delta v_{i}}{2 \sqrt{a_i^{max} b_i^{max}}}
\end{equation}
where $a_i^{max}$ is the maximum acceleration of $H_i$, $b_i^{max}$ is its maximum deceleration, $v_{0i}$ is its desired velocity, $s_{0i}$ is its desired following distance, $s_i$ is its current following distance, and $\Delta v_i$ is the current difference between its own velocity and the velocity of the preceding car.

For lateral control (i.e., lane-changing), we define the safety function for $H_i$ as
\begin{equation}
    z^S_i := p^F - p_{Hi} - s^{min}_i \geq 0 \land p_{Hi} - p^B - s^{min}_i \geq 0
\end{equation}
where $p^F$ is the position of the car in front of $H_i$ in the adjacent lane, $p^B$ is the position of the car behind $H_i$ in the adjacent lane, and $s^{min}_i$ is some minimum distance threshold desired by $H_i$. 
We define the incentive function for $H_i$ as
\begin{equation}
    z^I_i := v^F - v_{Hi} - \Delta v^{th}_i \geq 0
\end{equation}
where $v^F$ is the velocity of the car in front of $H_i$ in the adjacent lane, and $\Delta v^{th}_i$ is some minimum threshold desired by $H_i$ for the velocity difference between the two cars. 
Together, these two criteria encode that a human driver changes lanes if there is ample space in the adjacent lane and a potential increase in velocity following the lane change.
This is similar to real-world highway lane change behavior, where humans seek to increase their speed up to some desired value, as can be seen in IDM.

\subsection{Experimental Scenarios}
\begin{figure*}[t]
    \centering
    \subfloat[Illust. (\textbf{S1})]{\includegraphics[width=.15\linewidth, height=.1125\linewidth] {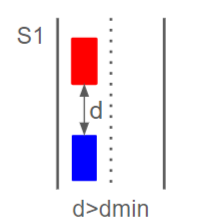}}
    \subfloat[Fol. dist. (\textbf{S1})]{\includegraphics[width=.15\linewidth] {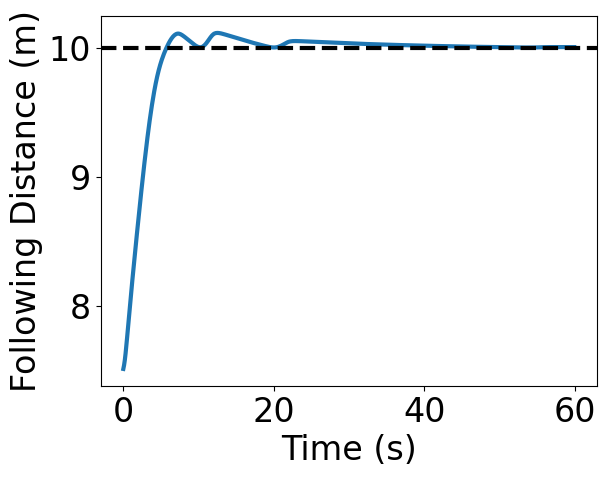}}
    \hfill
    \subfloat[Velocity (\textbf{S1})]{\includegraphics[width=.15\linewidth] {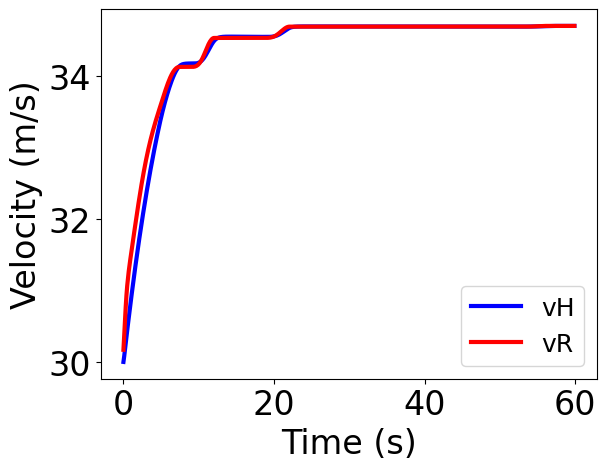}}
    \hfill\hspace{-20pt}
    \subfloat[Illust. (\textbf{S2})]{\includegraphics[width=.15\linewidth, height=.1125\linewidth] {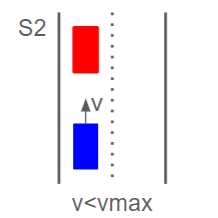}}
    \subfloat[Fol. dist. (\textbf{S2})]{\includegraphics[width=.15\linewidth] {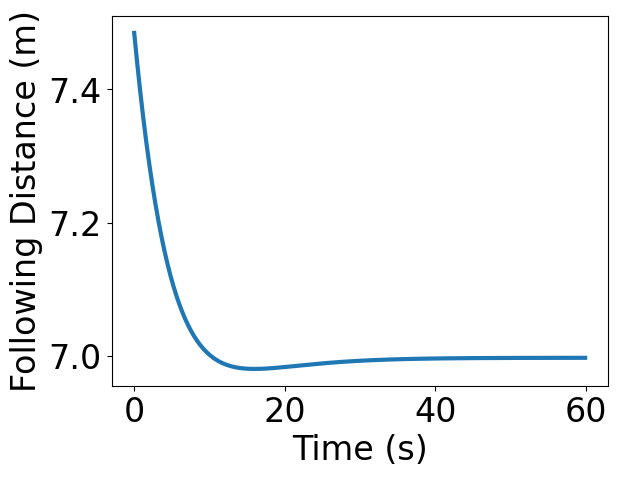}}
    \hfill
    \subfloat[Velocity (\textbf{S2})]{\includegraphics[width=.15\linewidth] {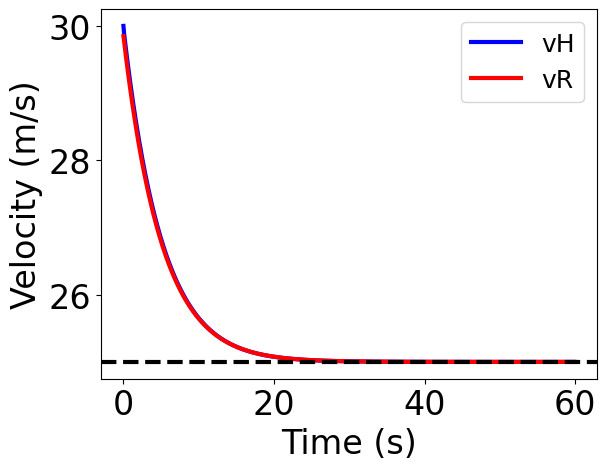}}
    
    \subfloat[Illust. (\textbf{S3})]{\includegraphics[width=.15\linewidth, height=.1125\linewidth] {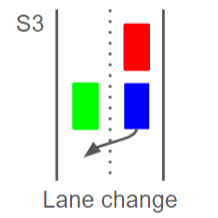}}
    \subfloat[Rel. pos. (\textbf{S3})]{\includegraphics[width=.15\linewidth] {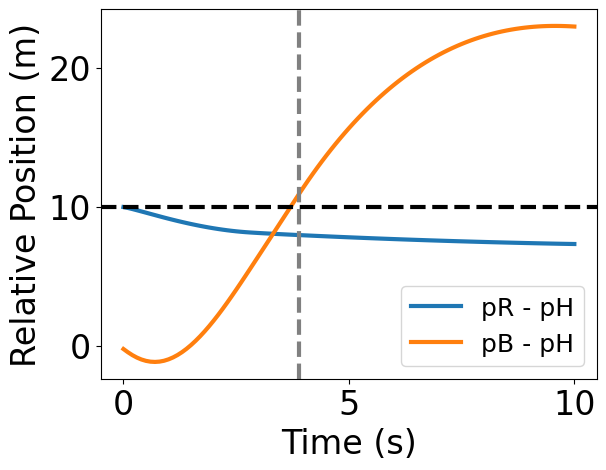}}
    \hfill
    \subfloat[Velocity (\textbf{S3})]{\includegraphics[width=.15\linewidth] {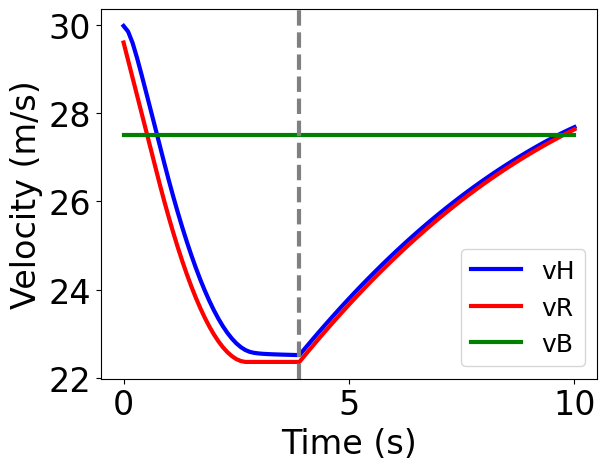}}
    \hfill\hspace{-20pt}
    \subfloat[Illust. (\textbf{SM1})]{\includegraphics[width=.15\linewidth, height=.1125\linewidth] {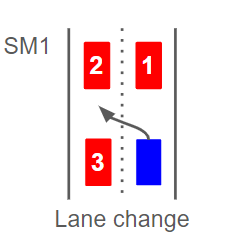}}
    \subfloat[Rel. pos. (\textbf{SM1})]{\includegraphics[width=.15\linewidth] {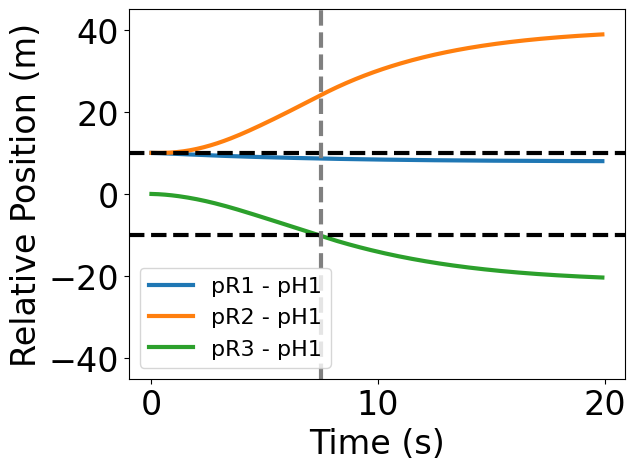}}
    \hfill
    \subfloat[Velocity (\textbf{SM1})]{\includegraphics[width=.15\linewidth] {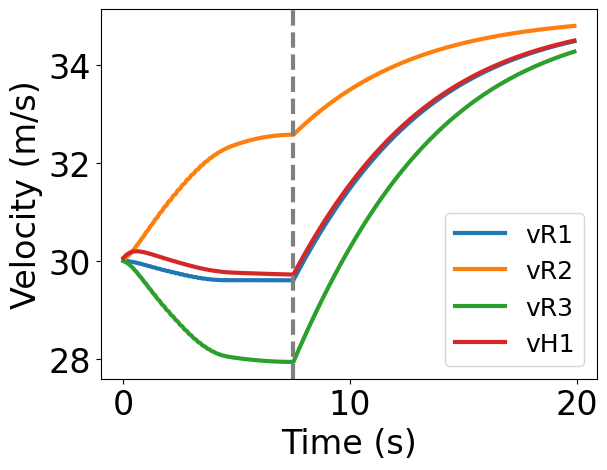}}
    
    \subfloat[Illust. (\textbf{SM2})]{\includegraphics[width=.15\linewidth, height=.1125\linewidth] {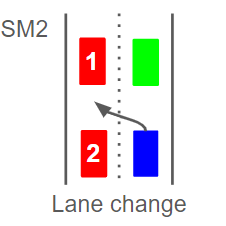}}
    \subfloat[Rel. pos. (\textbf{SM2})]{\includegraphics[width=.15\linewidth] {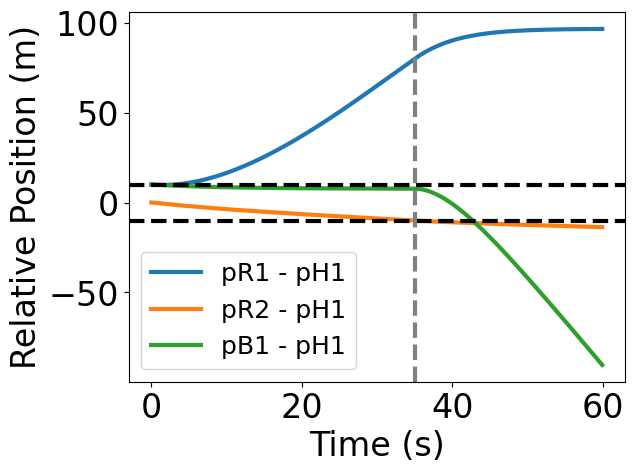}}
    \hfill
    \subfloat[Velocity (\textbf{SM2})]{\includegraphics[width=.15\linewidth] {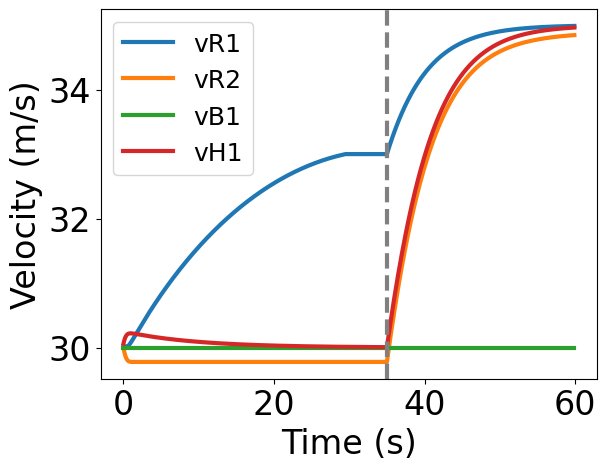}}
    \hfill\hspace{-20pt}
    \subfloat[Illust. (\textbf{SM3})]{\includegraphics[width=.15\linewidth, height=.1125\linewidth] {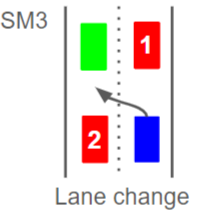}}
    \subfloat[Rel. pos. (\textbf{SM3})]{\includegraphics[width=.15\linewidth] {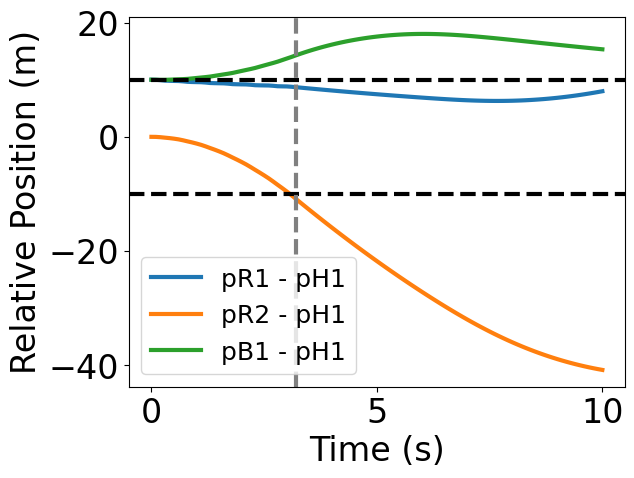}}
    \hfill
    \subfloat[Velocity (\textbf{SM3})]{\includegraphics[width=.15\linewidth] {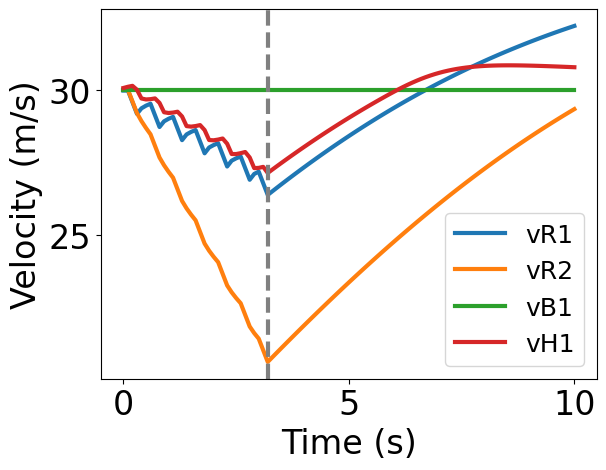}}
    \caption{Illustrations, human-relative position vs. time graphs, and velocity vs. time graphs for scenarios \textbf{S1}-\textbf{S3} and \textbf{SM1}-\textbf{SM3}. In illustrations, red boxes are robot cars, blue boxes are human-driven cars, and green boxes are background cars. In graphs, black dotted horizontal lines denote lower or upper bounds on the value via the influence objective, and gray dotted vertical lines denote the time at which a lane change occurs.}
    \label{fig:s_sm-graphs}
\end{figure*}

\begin{figure*}[t]
    \centering
    \subfloat[Illustration (\textbf{M1})]{\includegraphics[width=.225\linewidth, height=.16875\linewidth] {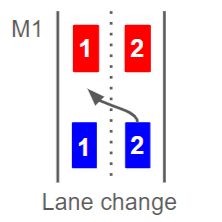}}
    \subfloat[$H_1$-relative position (\textbf{M1})]{\includegraphics[width=.225\linewidth] {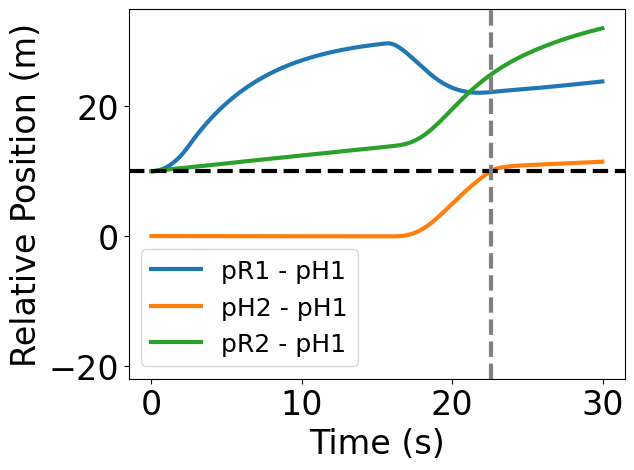}}
    \hfill
    \subfloat[$H_2$-relative position (\textbf{M1})]{\includegraphics[width=.225\linewidth] {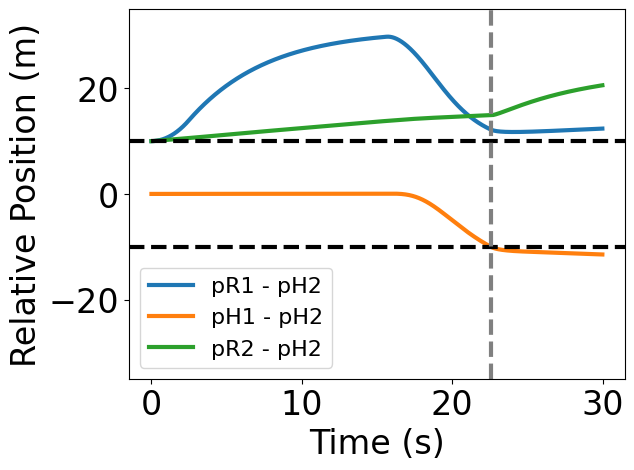}}
    \hfill
    \subfloat[Velocity (\textbf{M1})]{\includegraphics[width=.225\linewidth] {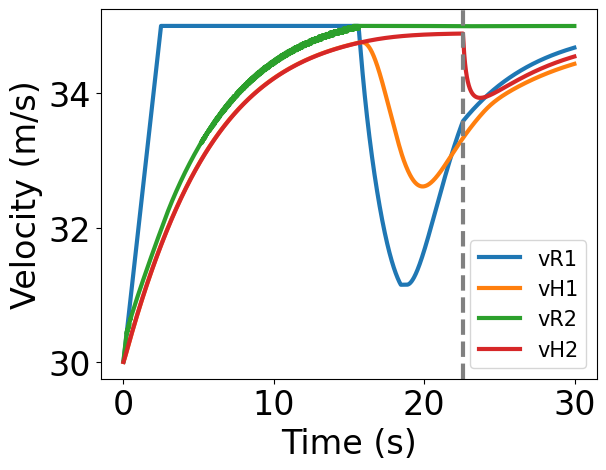}}
    
    \subfloat[Illustration (\textbf{M2})]{\includegraphics[width=.225\linewidth, height=.16875\linewidth] {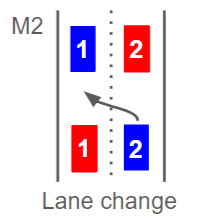}}
    \subfloat[$H_1$-relative position (\textbf{M2})]{\includegraphics[width=.225\linewidth] {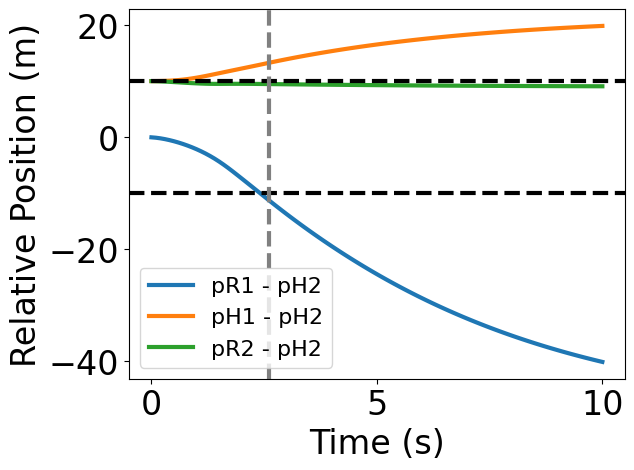}}
    \hfill
    \subfloat[$H_2$-relative position (\textbf{M2})]{\includegraphics[width=.225\linewidth] {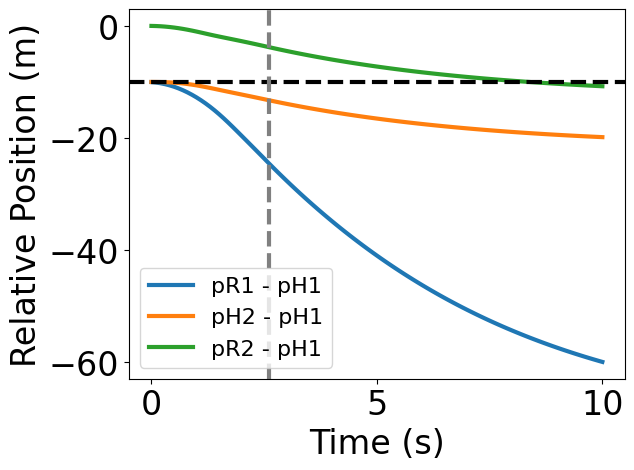}}
    \hfill
    \subfloat[Velocity (\textbf{M2})]{\includegraphics[width=.225\linewidth] {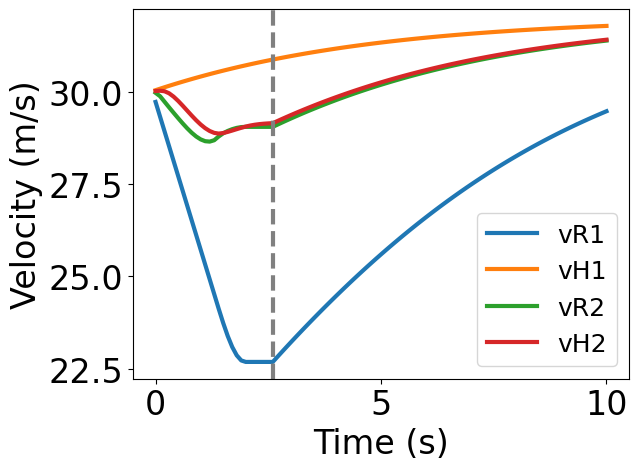}}
    
    \subfloat[Illustration (\textbf{M3})]{\includegraphics[width=.225\linewidth, height=.16875\linewidth] {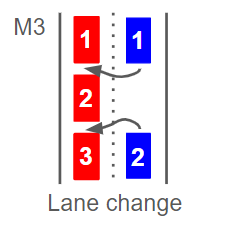}}
    \subfloat[$H_1$-relative position (\textbf{M3})]{\includegraphics[width=.225\linewidth] {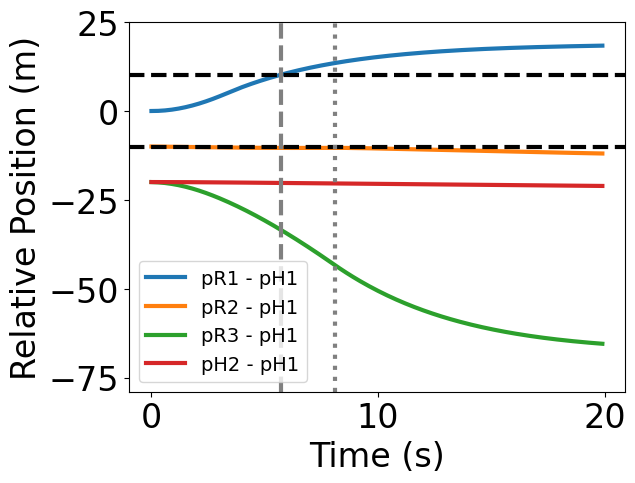}}
    \hfill
    \subfloat[$H_2$-relative position (\textbf{M3})]{\includegraphics[width=.225\linewidth] {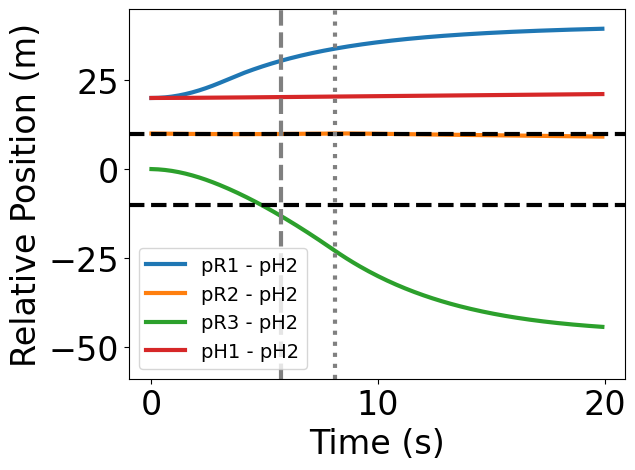}}
    \hfill
    \subfloat[Velocity (\textbf{M3})]{\includegraphics[width=.225\linewidth] {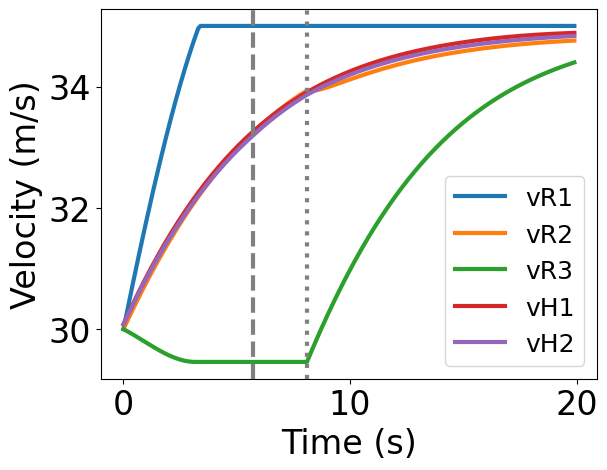}}
    
    \caption{Illustrations, human-relative position vs. time graphs, and velocity vs. time graphs for scenarios \textbf{M1}-\textbf{M3}. In illustrations, red boxes are robot cars, and blue boxes are human-driven cars. In graphs, black dotted horizontal lines denote lower or upper bounds on the value via the influence objective, and gray dotted vertical lines denote the time at which a lane change occurs.}
    \label{fig:m-graphs}

\end{figure*}

\begin{figure*}[t]
    \centering
    \subfloat[$t = 0$ \si{s}]{\label{fig:ex_a}\includegraphics[width=.25\linewidth]{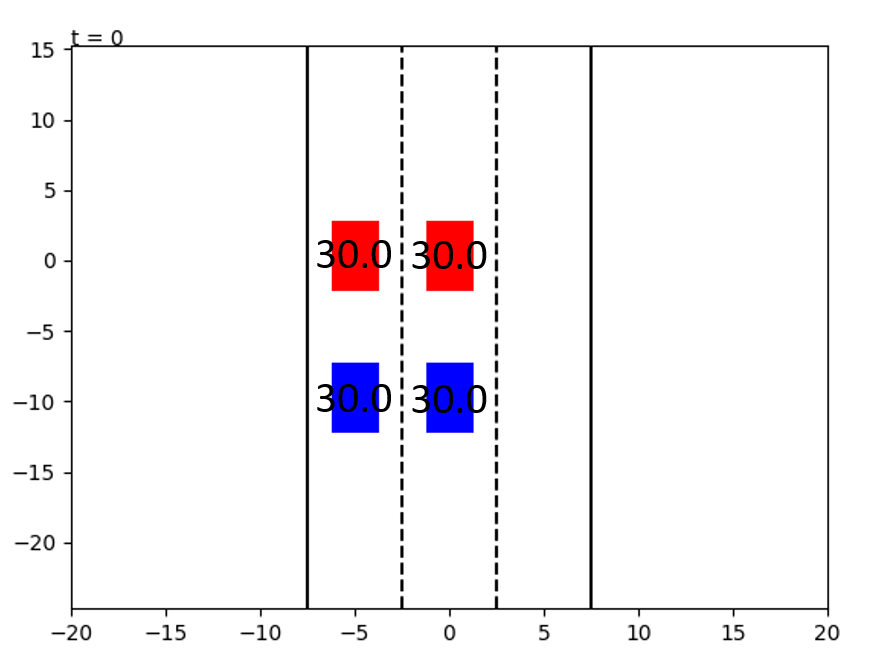}}
    \hfill
    \subfloat[$t = 10$ \si{s}]{\label{fig:ex_b}\includegraphics[width=.25\linewidth]{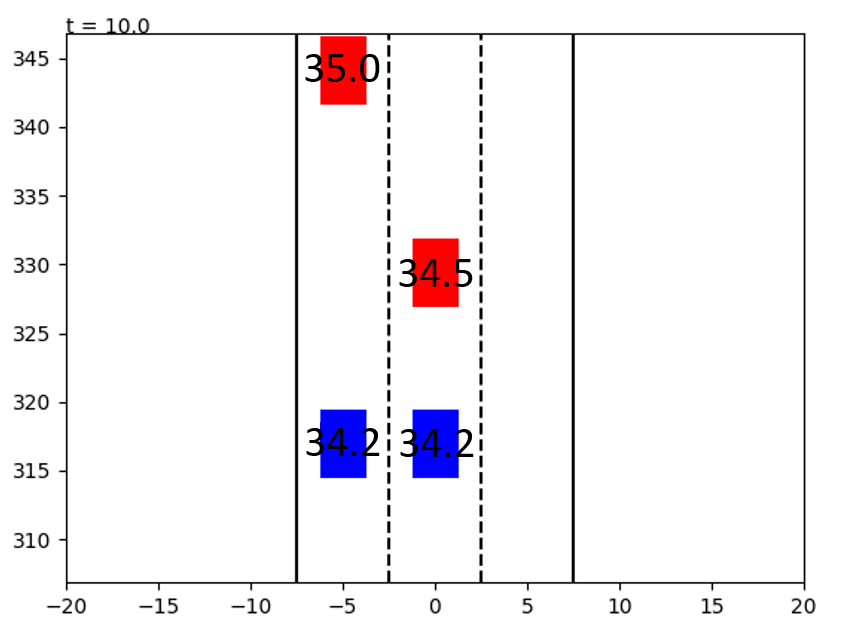}}
    \hfill
    \subfloat[$t = 20$ \si{s}]
    {\label{fig:ex_c}\includegraphics[width=.25\linewidth]{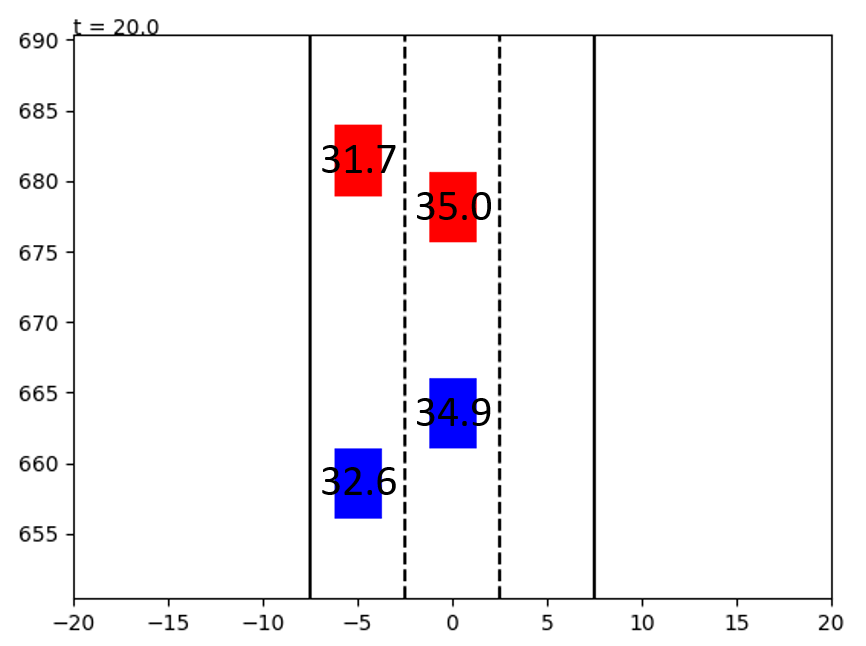}}
    \hfill
    \subfloat[$t = 30$ \si{s}] {\label{fig:ex_d}\includegraphics[width=.25\linewidth]{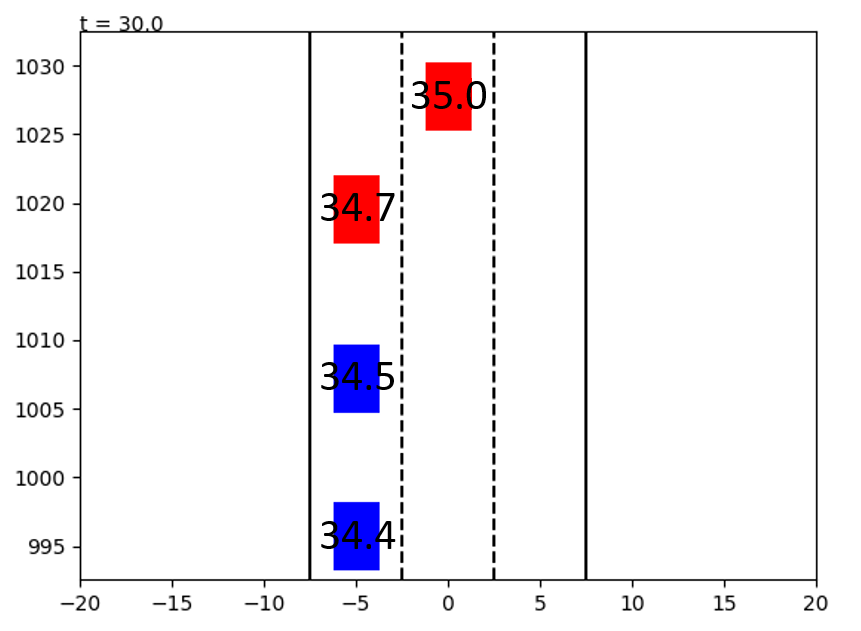}}
    \caption{Visualization of scenario \textbf{M1}, where two adjacent lanes each contain a robot car followed by a human-driven car, and the objective is to influence the human in the right lane to merge in between the two cars in the left lane. Red boxes are robot cars and blue boxes are human-driven cars. Overlaying each box is the car's velocity.}
    \label{fig:example}
\end{figure*}

Here we introduce nine low-level scenarios to demonstrate the feasibility of our framework under various objectives.
We examine three single-human single-robot scenarios (\textbf{S1}-\textbf{S3}), three single-human multi-robot scenarios (\textbf{SM1}-\textbf{SM3}), and three multi-human multi-robot scenarios (\textbf{M1}-\textbf{M3}).
For all scenarios, we have $\Delta t = 0.01$ \si{s}, car length $\ell = 5$ \si{m}, and each robot car $R_j$ has bounded velocity $v_{Rj} \in [0, 35]$ \si{m/s} and bounded acceleration $a_{Rj} \in [-4, 2]$ \si{m/s^2}.
For human behavior, we have $v_{0i} = 35$ \si{m/s}, $s^{min}_i = 10$ \si{m}, $\Delta v^{th}_i = 3$ \si{m/s} for all $H_i$, and humans follow the normal driving IDM parameters given in \cite{params}.
For simplicity, we omit the lane-change incentive criterion in scenarios \textbf{M1}-\textbf{M3}.
We use the cost function $J(\boldsymbol{u}) = \norm{\boldsymbol{u}}^2_2$ to minimize control effort in our optimization.
Fig. \ref{fig:s_sm-graphs} details \textbf{S1}-\textbf{S3} and \textbf{SM1}-\textbf{SM3} and their simulation results, Fig. \ref{fig:m-graphs} details \textbf{M1}-\textbf{M3} and their simulation results, and Fig. \ref{fig:example} visualizes \textbf{M1}.

\subsection{Example Derivation}
Here we provide an example derivation of the optimal control problem for scenario \textbf{M1}.
This includes the human driver lane change's forward and backward safety criteria.
First, we wish to enforce a minimum distance $s^{min}_2$ between $R_1$ and $H_2$ to allow for a lane change.
We express this as
\begin{equation}
    \psi_F := p_{R1} - p_{H2} - s^{min}_2.
\end{equation}
The time derivatives of $\psi_F$ are as follows:
\begin{align}
    \dot{\psi}_F &= \dot{p}_{R1} - \dot{p}_{H2} \nonumber \\
    \ddot{\psi}_F &= \ddot{p}_{R1} - \ddot{p}_{H2} = u_1 - f_2 \nonumber \\
    \dddot{\psi}_F &= \dot{u}_1 - \dot{f}_2 := \psi^h_F
\end{align}
Notice that $\ddot{\psi}_F$ contains $u_1$ but not $u_2$, so we take an additional time derivative of $\psi_F$ to obtain a $\dot{f}_2$ term in $\dddot{\psi}_F$.
This allows us to exploit the time derivative properties.
We apply \eqref{eq:fi-dot} to $\psi^h_F$ to obtain $\dot{u}_1 - \pdv{f_2}{v_{R2}} u_2 - \lambda_{22}$.
Intuitively, we have now translated human actions into robot controls.
Now, we just need to standardize everything to $\boldsymbol{u}$ terms rather than $\dot{\boldsymbol{u}}$, so we apply \eqref{eq:discrete-uj} to $\psi^h_F$ to obtain
\begin{equation}
    \frac{1}{\Delta t} u_1 - \pdv{f_2}{v_{R2}} u_2 - \frac{a_{R1}}{\Delta t} - \lambda_{22}.
\end{equation}
Finally, we have an explicit function of robot controls $u_1$ and $u_2$, so we can express our CBF constraint as
\begin{equation}
    g_F(\boldsymbol{x}, \boldsymbol{f}, \boldsymbol{u}) = \psi_F^h + \alpha^F_2 \ddot{\psi}_F + \alpha^F_1 \dot{\psi}_F + \alpha^F_0 \psi_F \geq 0.
\end{equation}

Next, we wish to enforce a minimum distance $s^{min}_2$ between $H_2$ and $H_1$ to allow for a lane change.
We express this as
\begin{equation}
    \psi_B := p_{H2} - p_{H1} - s^{min}_2.
\end{equation}
Notice that here, $\psi_B$ is not a function of the robot state.
The time derivatives of $\psi_B$ are as follows:
\begin{align}
    \dot{\psi}_B &= \dot{p}_{H2} - \dot{p}_{H1} \nonumber \\
    \ddot{\psi}_B &= \ddot{p}_{H2} - \ddot{p}_{H1} = f_2 - f_1 \nonumber \\
    \dddot{\psi}_B &= \dot{f}_2 - \dot{f}_1 := \psi_B^h
\end{align}
We reached $\dddot{\psi}_B$ to obtain $\dot{\boldsymbol{f}}$, so now we can again exploit the time derivative properties.
We apply \eqref{eq:fi-dot} to $\psi_B^h$ to obtain
\begin{equation}
    -\pdv{f_1}{v_{R1}} u_1 + \pdv{f_2}{v_{R2}} u_2 - \lambda_{11} + \lambda_{22}.
\end{equation}
We now have an explicit function of $u_1$ and $u_2$, so we can express our CBF constraint as
\begin{equation}
    g_B(\boldsymbol{x}, \boldsymbol{f}, \boldsymbol{u}) = \psi_B^h + \alpha_2^B \ddot{\psi}_B + \alpha_1^B \dot{\psi}_B + \alpha_0^B \psi_B \geq 0. 
\end{equation}

We have derived the linear constraints
\begin{equation}
\boldsymbol{g}(\boldsymbol{x}, \boldsymbol{f}, \boldsymbol{u}) :=
    \begin{bmatrix}
        g_F(\boldsymbol{x}, \boldsymbol{f}, \boldsymbol{u}) \\
        g_B(\boldsymbol{x}, \boldsymbol{f}, \boldsymbol{u})
    \end{bmatrix}
    \geq \boldsymbol{0}.
\end{equation}
We can formulate this as
\begin{equation}
    A^{M1} \boldsymbol{u} \geq \boldsymbol{b}^{M1}
\end{equation}
where
\begin{equation}
    A^{M1} := 
    \begin{bmatrix}
        (\alpha_2^F + \dfrac{1}{\Delta t}) & -\dfrac{\partial f_2}{\partial v_{R2}} \\[2ex]
        -\dfrac{\partial f_1}{\partial v_{R1}} & \dfrac{\partial f_2}{\partial v_{R2}} \\
    \end{bmatrix},~
    \boldsymbol{b}^{M1} := 
    \begin{bmatrix}
        \beta^F \\
        \beta^B
    \end{bmatrix}
    \nonumber
\end{equation}
with $\beta^F := \frac{a_{R1}}{\Delta t} + \lambda_{22} + \alpha_2^F f_2 - \alpha_1^F \dot{\psi}_F - \alpha_0^F \psi_F$ and $\beta^B := \lambda_{11} - \lambda_{22} - \alpha_2^B \ddot{\psi}_B - \alpha_1^B \dot{\psi}_B - \alpha_0^B \psi_B$.
Finally, we can pose the following QP to solve for the robot controls at each timestep:
\begin{align}
    \boldsymbol{u}^* = \operatorname*{argmin}_{\boldsymbol{u}} \hspace{5pt} \norm{\boldsymbol{u}}_2^2 ~~~ \text{s.t.} ~~ A^{M1} \boldsymbol{u} \geq \boldsymbol{b}^{M1}
\end{align}

\section{Case Studies}\label{cases}
In this section, we quantify the effectiveness of our framework and demonstrate its real-world applicability to two high-level objectives.

\subsection{Traffic Flow Optimization}
Here we consider a three-lane highway setting where each lane initially contains between 1 and 3 robot cars and between 1 and 3 human-driven cars.
The objective of the robot cars is to produce an increase in traffic flow by strategically influencing human-driven car lane changes.
In particular, we use one-dimensional $k$-means clustering on the humans' desired velocities to assign each car to a lane, where $k = 3$ and each resulting cluster represents a group of cars that should travel in the same lane.
The rationale for this lane assignment strategy is that the total traffic flow across a group of lanes is improved when the cars in each lane have similar desired velocities.
Once all human-driven cars have been assigned new lanes, each car is influenced to change lanes from its current lane to its assigned lane by the surrounding robot cars (up to $3$) using our optimal control framework.
Human-driven cars are selected individually for influence by order of index.
After all lane changes, each robot car follows IDM control with a desired velocity that matches the highest desired velocity of any human-driven car in its lane.

There is an equal likelihood of each number of robot cars and human-driven cars in each lane, and a given car has an equal likelihood of being a robot car and a human-driven car.
For each human-driven car $H_i$, we generate a desired velocity $v_{Hi} \sim \mathcal{U}(25, 40)$ and a lane-change velocity threshold $\Delta v^{th}_i \sim \mathcal{U}(2, 5)$.
We introduce a noise $\mathcal{N}(0, 0.2)$ to each human's desired following distance, $\mathcal{N}(0, 2)$ to the desired velocity, and $\mathcal{N}(0, 0.1)$ to the desired acceleration, deceleration, and time headway.
Each human's desired lane-change space threshold is equal to the desired following distance.
We introduce a noise $\mathcal{N}(0, 0.2)$ to the controls of the human-driven cars.

We use all cars' average velocity and the average difference between the humans' desired and actual velocities as metrics for traffic flow, where an increase in the former and a decrease in the latter indicate an improvement in traffic flow.
We simulated 100 trials of this scenario under these conditions, and compared the traffic flow performance prior to and following the human influence maneuvers, at $t = 0$ \si{s} and $t = 120$ \si{s}, respectively.
Fig. \ref{fig:flow-vel} shows the average velocity vs. time.
Dependent $t$-tests for paired samples showed a significant increase in average velocity ($t(100) = 3.829, p < 0.001$) and a significant decrease in the difference between desired and actual velocity ($t(100) = 7.146, p < 0.0001$) following the human influence maneuvers.
This implies that our framework is effective in improving mixed-autonomy highway traffic flow.

\subsection{Aggressive Behavior Mitigation}
Here we consider a single-lane highway setting where a robot car $R_1$ is preceded by a background car $B_1$ and followed by a human-driven car $H_1$ exhibiting aggressive following behavior.
This aggressive behavior is encoded as a close following distance, high velocity, and large acceleration and deceleration.
The robot car aims to mitigate this behavior by influencing a lower bound $s_{min}$ on the human-driven car's following distance, an upper bound $v_{max}$ on its velocity, and a lower bound $a_{min}$ and upper bound $a_{max}$ on its acceleration.
We formulate each of these intentions as CBF constraints using our framework.
We adopt the IDM parameters provided in \cite{params} to define the human's aggressive following behavior and the desired normal behavior.

We introduce a noise $\mathcal{N}(0, 0.2)$ to the human's desired following distance, $\mathcal{N}(0, 2)$ to its desired velocity, and $\mathcal{N}(0, 0.1)$ to its desired acceleration, deceleration, and time headway.
We also introduce a noise $\mathcal{N}(0, 0.2)$ to the controls of the human and background cars.
We fix the initial robot car position $p_{R1} = 0$, and generate an initial human-driven car position $p_{H1} \sim \mathcal{U}(-3 \ell, - \ell)$ and background car position $p_{B1} \sim \mathcal{U}(\ell, 3 \ell)$.
We also generate initial velocities $v_{H1}, v_{R1}, v_{B1} \sim \mathcal{U}(25, 35)$.

\begin{figure}[t]
    \centering
    \begin{minipage}{.475\columnwidth}
        \includegraphics[width=\linewidth,height=0.7\linewidth]{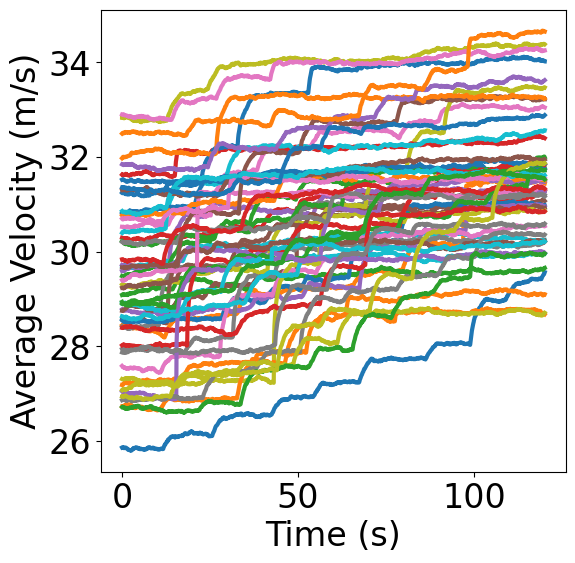}
        \caption{Average velocity vs. time for traffic flow optimization.}
        \label{fig:flow-vel}
    \end{minipage}\hfill
    \begin{minipage}{.475\columnwidth}
        \includegraphics[width=\linewidth]{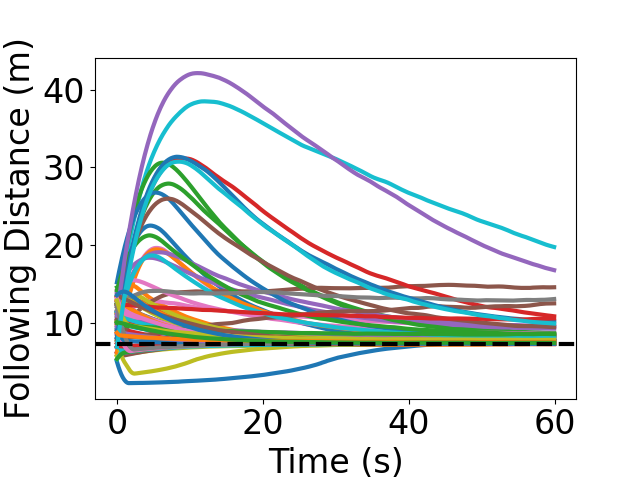}
        \caption{Human following distance vs. time for aggression mitigation.}
        \label{fig:agg-dist}
    \end{minipage}
    \begin{minipage}{.475\columnwidth}
        \includegraphics[width=\linewidth]{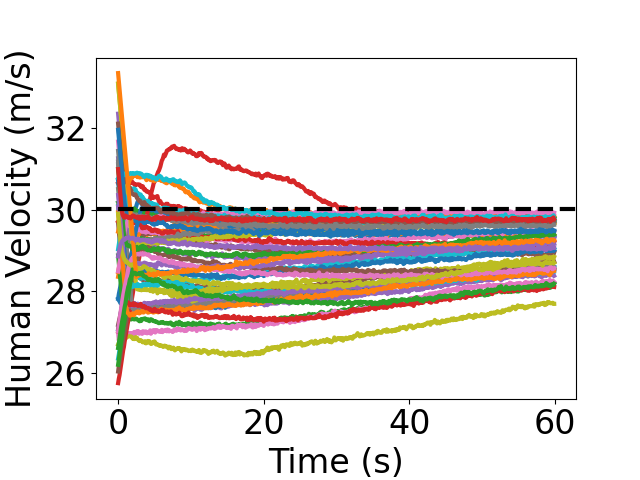}
        \caption{Human velocity vs. time for aggression mitigation.}
        \label{fig:agg-vel}
    \end{minipage}\hfill
    \begin{minipage}{.475\columnwidth}
        \includegraphics[width=\linewidth]{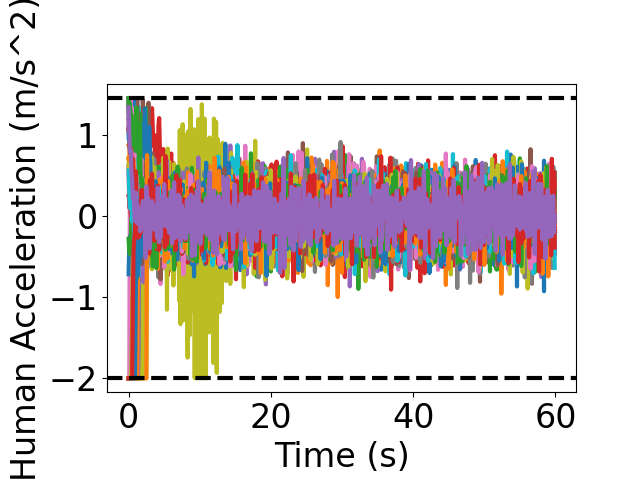}
        \caption{Human acceleration vs. time for aggression mitigation.}
        \label{fig:agg-acc}
    \end{minipage}
    \vspace{-10pt}
\end{figure}

We use the average magnitude of jerk as a metric for aggressive following behavior.
Higher values of average jerk magnitude were shown in \cite{jerk} to correlate with higher levels of aggression.
We simulated 100 trials of this scenario under these conditions, and compared the average magnitude of jerk from $t = 0$ \si{s} to $t = 60$ \si{s} with that of a control condition where the robot car instead follows IDM control with normal driving parameters.
Fig. \ref{fig:agg-dist} shows human following distance vs. time, Fig. \ref{fig:agg-vel} shows human velocity vs. time, and Fig. \ref{fig:agg-acc} shows human acceleration vs. time.
A dependent $t$-test for paired samples showed a significant decrease in average jerk magnitude when human influence was employed ($t(100) = 2.368, p < 0.01$).
This implies that our framework is effective in mitigating aggressive human following behavior in highway settings.

\section{Conclusion}\label{conclusion}
In this work, we presented a novel optimal control framework for influencing human driving behavior in mixed-autonomy traffic.
We leveraged control barrier functions to formulate the problem of human influence as a constrained optimization problem on the controls of the surrounding robot cars.
We demonstrated our framework's applicability to various objectives and configurations, including multi-robot and multi-human scenarios.
We validated its effectiveness in the two real-world objectives of traffic flow optimization and aggressive behavior mitigation.
Through our framework, we advance the state of the art by contributing superior versatility in the autonomous influence of human driving behavior.
In future work, we intend to expand our framework's compatibility with various human behavior models and further study its flexibility and scalability.

\end{document}